\documentclass[useAMS,usenatbib]{mn2e}

\usepackage{graphicx}
\usepackage{float}

\title[Detectability of nebular lines]
{Nebular line emission from $z > 7$ galaxies in a cosmological simulation: rest-frame UV to Optical lines }
\author[Shimizu et al.]
{Ikkoh Shimizu$^{1,2}$\thanks{E-mail: shimizu@vega.ess.sci.osaka-u.ac.jp}, 
Akio K. Inoue$^{3}$, Takashi Okamoto$^{4}$ and Naoki Yoshida$^{5,6}$\\
$^{1}$Department of Astronomy, School of Science, The University of Tokyo, 7-3-1 Hongo, Bunkyo-ku, Tokyo 113-0033 \\
$^{2}$Department of Earth \& Space Science, Graduate School of Science, Osaka University, 1-1 Machikaneyama, Toyonaka, Osaka 560-0043, Japan \\
$^{3}$College of General Education, Osaka Sangyo University, 3-1-1 Nakagaito, Daito, Osaka 574-8530, Japan \\
$^{4}$Department of Cosmosciences Graduate School of Science, Hokkaido University, N10 W8, Kitaku, Sapporo, 060-0810, Japan \\
$^{5}$Department of Physics, The University of Tokyo, 7-3-1 Hongo, Bunkyo-ku, Tokyo 113-0033, Japan \\
$^{6}$Kavli Institute for the Physics and Mathematics of the Universe, TODIAS, \\
The University of Tokyo, 5-1-5 Kashiwanoha, Kashiwa, Chiba 277-8583, Japan}
\begin{document}

\date{In original form 2014 January 4}

\pagerange{\pageref{firstpage}--\pageref{lastpage}} \pubyear{2013}

\maketitle

\label{firstpage}

\begin{abstract}
We have performed very large and high resolution cosmological hydrodynamic simulations in order to investigate detectability of nebular lines in the rest-frame ultraviolet (UV) to optical wavelength range from galaxies at $z > 7$. 
We find that the expected line fluxes are very well correlated with apparent UV magnitudes. 
The C {\sc iv} $1549 \rm \AA$ and the C {\sc iii}] $1909 \rm \AA$ lines of galaxies brighter than $26~ \rm AB$ are detectable with current facilities such as the VLT/XShooter and the Keck/MOSFIRE. 
Metal lines such as C {\sc iv} $1549 \rm \AA$, C {\sc iii}] $1909 \rm \AA$, [O {\sc ii}] $3727 \rm \AA$ and [O {\sc iii}] $4959/5007 \rm \AA$ can be good targets for the spectroscopic observation with Thirty Meter Telescope (TMT), European Extremely Large Telescope (E-ELT), Giant Magellan Telescope (GMT) and James Webb Space Telescope (JWST). 
H$\alpha$ and H$\beta$ lines are also expected to be detectable with these telescopes.  
We also predict detectability of the nebular lines for $z > 10$ galaxies that will be found with JWST, Wide-Field Infrared Survey Telescope (WFIRST) and First Light And Reionization Explorer (FLARE) $(11 \leq z \leq 15$). 
We conclude that the C {\sc iv} $1549 \rm \AA$, C {\sc iii}] $1909 \rm \AA$, [O {\sc iii}] $4959/5007 \rm \AA$ and H$\beta$ lines even from $z \sim 15$ galaxies can be strong targets for TMT, ELT and JWST. 
We also find that the magnification by gravitational lensing is of great help to detect such high-$z$ galaxies. 
According to our model, the C {\sc iii}] $1909 \rm \AA$ line in $z > 9$ galaxy candidates is detectable even using the current facilities. 
\end{abstract}


\section{Introduction}
Understanding physical properties of the high-$z$ galaxies beyond $z = 7$ is very important, 
because such galaxies are not only building blocks of lower-$z$ galaxies and local galaxies but also the leading candidate of ionizing sources for the cosmic reionization. 
Such galaxies are much fainter objects because they are very distant objects and less massive systems. 
Thus, in order to observe these galaxies, ultra deep surveys are necessary. 
The Wide Field Camera (WFC3) installed on the Hubble Space Telescope (HST) in 2009 have detected many $z > 7$ galaxy candidates using the dropout (Lyman break) technique 
\citep{Yan2011, Ellis2013, McLure2013, Schenker2013, Dunlop2013, Robertson2013, Ono2013, Koekemoer2013, Oesch2013, Bouwens2014a, Finkelstein2014, Bouwens2015}. 
The combination of gravitational lensing by foreground galaxy clusters and the WFC3 camera have also discovered even intrinsically fainter high-$z$ galaxy candidates \citep{Oesch2014, Ishigaki2015}. 

Many spectroscopic surveys for the Ly$\alpha$ line of $z > 7$ galaxies have been attempted to confirm spectroscopic redshift of these galaxies 
\citep[e.g.,][]{Pentericci2011, Pentericci2015, Caruana2012, Caruana2014, Schenker2012, Shibuya2012, Finkelstein2013, Treu2013, Schenker2014, Vanzella2014, Oesch2015}. 
Now, the highest confirmed redshift using Ly$\alpha$ line is $z = 8.68$ \citep{Zitrin2015b} which is the first and only example of a successful Ly$\alpha$ line observation at $z > 8$. 
Other observations so far have not succeeded in the Ly$\alpha$ line detection at $z > 9$ \citep{Brammer2013, Bunker2013, Capak2013, Matthee2014}. 
Moreover, the spectroscopic surveys for Ly$\alpha$ line revealed that the fraction of high-$z$ galaxies with detectable Ly$\alpha$ emission decreases from $z = 6$ \citep[e.g.,][]{Stark2011, Ono2012, Tilvi2014}. 
These observational results imply that Ly$\alpha$ attenuation by the neutral hydrogen in the intergalactic medium (IGM) is significant 
because the neutral fraction of the hydrogen is high at such higher-$z$ before the completion of the cosmic reionization. 
The Ly$\alpha$ line detection at $z > 8$ may be very difficult even by future telescopes such as the James Webb Space Telescope (JWST), the European Extremely Large Telescope (E-ELT), the Giant Magellan Telescope (GMT) and the Thirty Meter Telescope (TMT). 
Thus, it appears to be important to consider the detectability of alternative lines instead of the Ly$\alpha$ line. 
Furthermore, the detection of high-$z$ metal lines helps to understand when and how the metal enrichment proceeds at very early Universe.  

The surveys for the far-infrared (FIR) lines using the Atacama Large Millimetre/submillimetre Array (ALMA) can be very powerful tool to confirm redshifts of very high-$z$ galaxies. 
The [C {\sc ii}] $158 \rm \mu m$ line is a well known brightest line in the FIR region. 
The line has been detected in many QSOs and submm galaxies (SMGs) at various redshifts \citep{Maiolino2005, Iono2006, Capak2011, Vieira2013, Wang2013}. 
[C {\sc ii}] $158 \rm \mu m$ line in some high-$z$ LBGs has also been discovered \citep{Capak2015}. 
Interestingly, almost all survey have failed to detect [C {\sc ii}] $158 \rm \mu m$ line in high-$z$ Lyman $\alpha$ emitters (LAEs) \citep{Walter2012, Kaneker2013, Ouchi2013, Ota2014, Watson2015}. 
This fact disadvantages to explore high-$z$ [C {\sc ii}] $158 \rm \mu m$ line because the fraction of young galaxies like LAEs increases with redshift. 
The [C {\sc ii}] $158 \rm \mu m$ is thought to be originated mainly from photo-dissociation regions (PDRs), the modelling of which is very complicated. 
This implies that the reliable theoretical prediction for the [C {\sc ii}] $158 \rm \mu m$ line is very difficult.  
In our previous work \citep{Inoue2014a}, we concluded that the [O {\sc iii}] $88 \rm \mu m$ line from H {\sc ii} regions is potentially the best line to measure redshifts of galaxies in the early Universe. 
The modelling of this line is simpler than that of the [C {\sc ii}] $158 \rm \mu m$ line and the ALMA band 7 covers the line at $8.1 < z < 11.3$. 

There are also various strong lines in the rest-frame UV to optical region. 
In the High-redshift Emission Line Survey (HiZELS) with narrow-band filters in the $J$, $H$ and $K$ bands, 
many H$\alpha$, [O {\sc ii}] $3727 \rm \AA$ and [O {\sc iii}] $5007 \rm \AA$ emitters have been discovered at $0.4 < z < 2.2$ \citep{Geach2008, Sobral2009, Sobral2014, Sobral2015a}. 
The spectroscopic surveys have also been performed and discovered many lines such as C {\sc iii}] $1909 \rm \AA$, [O {\sc ii}] $3727 \rm \AA$, [O {\sc iii}] $5007 \rm \AA$ and H$\alpha$ in the UV to the optical region at $z < 3$ \citep{Silverman2014, Kashino2014, Nakajima2014, Stark2014, Sanders2015}. 
\citet{Stark2015a} detected C {\sc iii}] $1909 \rm \AA$ emission in two high-$z$ galaxies at $z = 6.029$ and $7.213$. 
The C {\sc iii}] $1909 \rm \AA$ emission in unusually luminous galaxy at $z = 7.730$ have been detected by \citet{Stark2016}. 
In addition, \citet{Stark2015b} discovered the C {\sc iv}] $1549 \rm \AA$ and O {\sc iii}] $1665 \rm \AA$ emission lines in $z = 7.045$ galaxy confirmed by the Ly$\alpha$ emission line. 
More recently, a strong He {\sc ii} $1640 \rm \AA$ was found in $z = 6.6$ LAEs \citep[CR7,][]{Sobral2015b}. 
Interestingly, the CR7 has no metal lines. 
\citet{Sobral2015b} suggested the existence of metal-free stars so-called Pop III stars and/or the formation of a direct collapse black hole. 
Now, the source of strong He {\sc ii} $1640 \rm \AA$ in the CR7 is hotly debated. 
\citep[e.g.,][]{Agarwal2015, Hartwig2015, Pacucci2015, Pallottini2015}.

The UV to optical lines are very useful for studying the physical properties of galaxies such as the gas metallicity, the radiation filed strength and the gas density \citep[e.g.,][]{Nagao2006, Mannucci2010, Nakajima2014, Sanders2015}. 
Nevertheless, there have been only a few line surveys at higher redshift ($z > 6$) using the ground-based telescopes due to the observational difficulty. 
This is because the wavelength of these lines in the high-$z$ Universe is redshifted into the IR region. 
The next generation telescopes such as the JWST, the E-ELT, the GMT and the TMT can perform a deeper survey than the current ones in the IR regime. 
Thus, the metal lines can be good targets for these future telescopes and useful to diagnose the physical properties of very high-$z$ galaxies ($z > 7$). 
Unfortunately, only small number of bright galaxies at $z > 7$ appropriate for the follow-up survey can be discovered by these telescopes due to their narrow field-of-view (FOV) ($\sim 30~\rm arcmin^2$). 
The Wide-Field Infrared Survey Telescope (WFIRST) and the First Light And Reionization Explorer (FLARE), which are also future telescopes, are the IR deep and wide-field imaging surveyors. 
The FOV of these telescopes are much larger than the TMT, the E-ELT, or the JWST. 
Many bright high-$z$ galaxies or very rare objects, which are good targets for spectroscopic follow-up with the TMT, the E-ELT, or the JWST, will be discovered.

Many theoretical studies have been made on very high-$z$ galaxies \citep[e.g.,][]{Finlator2011, Dayal2013, Shimizu2014, Clay2015, Feng2016}. 
In these papers, they investigated the physical properties of star-forming galaxies at high-$z$ 
such as the halo mass, the stellar mass and the UV magnitude. 
They also revealed the evolution of the star-formation density, the stellar mass functions (SMFs), the UV luminosity function (LFs) and the metal enrichment in the IGM. 
There are some studies for high-$z$ galaxies with nebular emission consideration. 
\citet{Schaerer2009} studied nebular emission effects in the spectral energy distribution (SED). 
They showed that the estimated stellar mass, the age and the extinction strength change in the cases of with or without the nebular emission. 
\citet{Inoue2011} performed galaxy SED calculations in the UV-to-optical wavelength range with the nebular emission using the photo-ionisation code {\scriptsize CLOUDY} \citep{cloudy}. 
In the paper, they discussed criteria of equivalent widths of some lines and broad-band colours to select extremely metal-poor and metal-free galaxies. 
\citet{Wilkins2013} took into account the effect of the nebular emission on observational colour selections using a large cosmological hydrodynamical simulation.  
They claimed that the photometric redshift estimation is strongly affected by the strength of the nebular emission. 
Although there are many studies for very high-$z$ galaxies with the nebular emission, 
studies about the emission line detectability in these high-$z$ galaxies are few. 
 
In this study, we examine whether some emission lines in the rest-frame UV to optical wavelength of galaxies at $z = 7$--$10$ are detectable with the current and future telescopes. 
Next, we discuss the detectability of the lines at even $z > 10$ in the future surveys. 

In section 2, we describe our numerical simulations and calibrate the parameters therein. 
We virtually observe the galaxies located in our simulation through a light-cone output and select simulated galaxies by applying a relevant magnitude limit. 
In section 3, we present some line fluxes in the rest-frame UV to optical wavelength of selected galaxies in our simulation.
In section 4, we make predictions using the FLARE, the WFIRST, JWST and TMT telescopes. 
The final section is devoted to our conclusion.

Throughout this paper, we adopt a $\Lambda$CDM cosmology 
with the matter density $\Omega_{\rm{M}} = 0.3175$, 
the cosmological constant $\Omega_{\Lambda} = 0.6825$, 
the Hubble constant $h = 0.6711$ in the unit of $H_0 = 100 {\rm ~km ~s^{-1} ~Mpc^{-1}}$ and 
the baryon density $\Omega_{\rm B} = 0.04899$. 
The matter density fluctuations are normalised by setting
$\sigma_8 = 0.8344$ \citep{Planck}. 
All magnitudes are quoted in the AB system \citep{Oke1990}. 
The assumed initial mass function (IMF) in the observational data and in our simulation 
is always the Chabrier IMF with the mass range of 0.1--100 $\rm M_\odot$ \citep{Chabrier2003}.

\section{Theoretical Model}
In this section, we describe our cosmological hydrodynamic simulation, calculation of the spectral energy distribution (SED) of the simulated galaxies 
and the treatment of the dust attenuation for these galaxies.

\subsection{The Hydrodynamic Simulation}
We performed high-resolution numerical simulations with an updated version of the Tree-PM smoothed particle hydrodynamics (SPH) code {\scriptsize GADGET-3} 
which is a successor of Tree-PM SPH code {\scriptsize GADGET-2} \citep{Gadget}. 
We implemented relevant physical processes to galaxy formation such as the star formation, the supernova (SN) feedback 
and the chemical enrichment following papers \citep{Okamoto2008, Okamoto2009, Okamoto2010}. 
In our previous work \citep{Okamoto2014}, the radiation pressure and the AGN feedback were newly implemented. 
We consequently reproduced various observational quantities such as the stellar mass functions (SMFs), the cosmic star formation history, 
the galaxy downsizing, and the relation between stellar mass and metallicity from $z = 4$ to $z = 0$. 
The details of these processes are found in the above references. 

Here, we briefly introduce settings of our simulations. 
We employ a total of $2 \times 1280^3$ particles for dark matter and gas in a comoving volume of $50 h^{-1}{\rm ~Mpc}$ cube.
The mass of a dark matter particle is $4.44 \times 10^6 h^{-1}{\rm M_{\odot}}$ 
and that of a gas particle is initially $8.11 \times 10^5 h^{-1}~{\rm M_{\odot}}$, respectively. 
The softening length for the gravitational force is set to be $2~ h^{-1}{\rm kpc}$ in comoving unit. 
The SPH gas particles can spawn star particles when they satisfies a set of standard criteria for star formation, and then, their mass is reduced. 
SPH particles around evolved star particles get the mass including metal elements due to SNe and the stellar mass loss in the AGB phase, resulting in the mass reduction of the star particles. 
In order to identify simulated galaxies, 
we run a friends-of-friends (FoF) group finder with a comoving linking length of 0.2 in units of the mean particle separation 
to identify groups of dark matter particles as haloes. 
Then, we find gravitationally bound groups of at least 32 total (dark matter + SPH + star) particles 
as substructures (subhaloes) in each FoF group using SUBFIND algorithm developed by \citet{Springel2001}. 
We regard substructures that contain at least 10 star particles as our simulated galaxies. 
We note that the lowest dark halo and stellar masses are around $10^8~\rm M_\odot$ and $10^6~\rm M_\odot$, respectively. 

\begin{figure}
\includegraphics[width = 80mm]{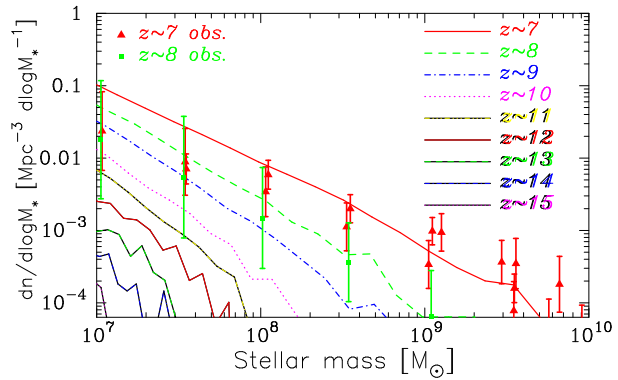}
\caption{The stellar mass functions (SMF) at $z = 7$ and higher. 
The line styles and their corresponding redshifts are noted in the panel. 
For $z = 7$ and $8$, we also plot observational data as the points with error bars taken from \citet{Gonzalez2011}, \citet{Grazian2015} and \citet{Song2015}. } 
\label{SMF}
\end{figure}

In order to calibrate parameters in the code such as the SN feedback efficiency, 
we compare our model with the observed stellar mass function (SMF) of galaxies. 
Fig. \ref{SMF} represents our SMF at $z = 7$ to $z = 15$.  
We also plot the observational results (the points with error bars) at $z = 7$ and $z = 8$ taken from \citet{Gonzalez2011}, \citet{Grazian2015} and  \citet{Song2015}. 
Our model reproduces the observed SMF within error-bars. 
At the massive end of SMFs ($> 10^{10}~ \rm M_{\rm \odot}$), our model may underestimate the SMF because our simulation box is not sufficiently large. 
On the other hand, our simulation tends to overproduce less massive galaxies ($< 10^{7.5}~ \rm M_{\odot}$) when comparing the observational data at $z \sim 7$. 
This may indicate that the SFR of less massive galaxies in our simulation is $\sim 2$ times higher than real ones. 
In our previous work \citep{Shimizu2013}, 
we argue that the shape of the SMF depends on the definition of the stellar mass, 
i.e., with or without the remnant mass as the stellar mass. 
The effect is important for low-$z$ galaxies, 
but, it is negligible at the high-$z$ discussed in this paper. 

\subsection{SED Calculation of Simulated Galaxies}
After the identification of the simulated galaxies, we calculate their SEDs. 
The SED of each star particle, which has its own age, metallicity, and mass, is calculated by using the population synthesis code {\scriptsize P\'{E}GASE2} \citep{PEGASE}. 
Then, we sum up the SEDs of the star particles composing a simulated galaxy to obtain the total intrinsic SED of the galaxy. 
We also adopt the nebular continuum and the Hydrogen recombination lines (Ly$\alpha$, H$\alpha$, H$\beta$, and H$\gamma$) calculated by {\scriptsize P\'{E}GASE2} with the Case B approximation. 
Moreover, the metal lines emitting from H {\sc ii} regions are calculated based on \citet{Inoue2011} and \citet{Inoue2014a}, 
in which expected line fluxes are calculated by the photo-ionization code {\scriptsize CLOUDY} \citep{cloudy}. 
These line luminosity ($L_{\rm line}$) are proportional to H$\beta$ luminosity ($L_{\rm H\beta}$), 
\begin{equation}
L_{\rm line} = (1 - f_{\rm esc}) C_{\rm line}(Z) L_{\rm H\beta}, 
\end{equation}
where $f_{\rm esc}$ is the Lyman continuum escape fraction and $C_{\rm line}(Z)$ is the metallicity dependent emission efficiency for each line which is calculated in \citet{Inoue2011} and \citet{Inoue2014a}, respectively. 
Measurements of $f_{\rm esc}$ of an order of 0.01--0.1 have been obtained at $z \sim3$ \citep[e.g.,][]{Inoue2005, Iwata2009} and $f_{\rm esc} \sim 0.2$ at $z > 5$ has been favoured for  cosmic reionization \citep{Inoue2006, Shapley2006, Siana2015}. 
However, we simply assume $f_{\rm esc} = 0$ in this paper, yielding the maximum fluxes of nebular emission lines. 
It is worth describing effect of a different choice of stellar population synthesis (SPS) model. 
One of important point for line emissivity calculation is the Lyman continuum photon production efficiency per nonionizing UV ($\sim 1500~ \rm \AA$)
because line emissivity strongly correlates with the efficiency. 
The difference between {\scriptsize P\'{E}GASE2}, {\scriptsize Starburst99} \citep{Leitherer1999}, {\scriptsize Maraston} \citep{Maraston2005} and {\scriptsize BC03} \citep{Bruzual2003} is very small \citep[e.g.,][]{Inoue2014a, Wilkins2016}. 
However, the efficiency of {\scriptsize BPASS/binary} \citep{Stanway2015} is about a factor of 2 higher than others \citep[see also][]{Wilkins2016}). 
Hence, uncertainty in the line flux estimation due to the choice of SPS models is also a factor of 2. 

\begin{figure}
\includegraphics[width = 80mm]{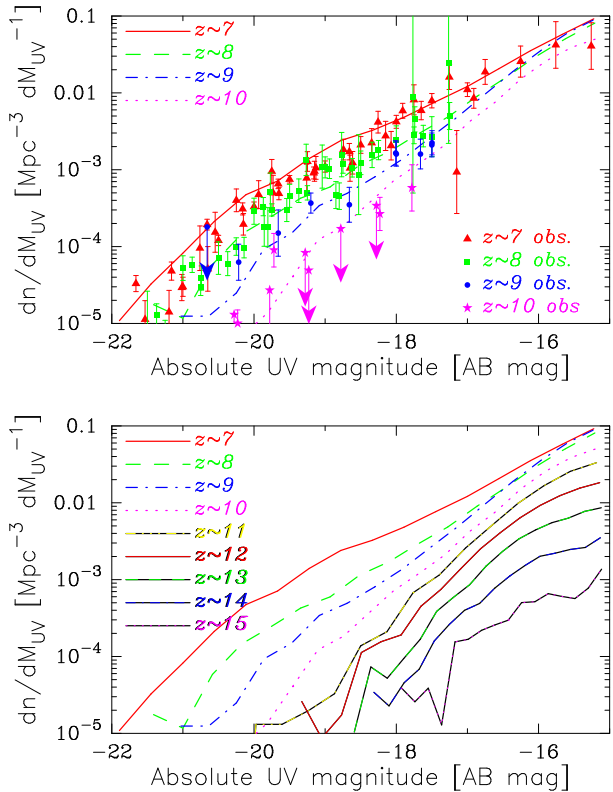}
\caption{The UV luminosity functions. The top panel represents the UV LFs from $z = 7$ to $z = 10$. 
The bottom panel represents the UV LFs from $z = 7$ to $z = 15$. 
The line styles and their corresponding redshifts are noted in the panel. 
The points with error-bars or arrows in the left panel are shown the observational data 
\citep{Robertson2010, Bouwens2011a, Bouwens2011b, Bradley2012, Oesch2012, McLure2013, Oesch2013, Schenker2013, Bouwens2014a, Bouwens2015, Atek2015, McLeod2016}. } 
\label{UVLF}
\end{figure}

After calculating the intrinsic SEDs of simulated galaxies, then, we consider the dust attenuation for the continuum and the lines of them. 
The procedure of dust attenuation for the continuum is same as our previous work \citep{Shimizu2014}. 
We adopt the Calzetti law \citep{Calzetti2000} for the shape of the attenuation law.  For the attenuation amplitude, 
we calculate the escape probability of UV photons at $1500~{\rm \AA}$ ($f_{\rm UV}^{\rm cont}$). 
We apply the sandwich model for $f_{\rm UV}^{\rm cont}$ \citep{Shimizu2014, Xu1995}, 
\begin{equation}
f_{\rm UV}^{\rm cont} = \frac{1 - \delta}{2} (1 + {\rm e}^{- \tau_{\rm d}}) + \frac{\delta}{\tau_{\rm d}} (1 - {\rm e}^{- \tau_{\rm d}}),  
\label{eq:ep}
\end{equation}
where $\delta$ is a parameter whose value is from $0$ to $1$ and $\tau_{\rm d}$ is the UV (1500$\AA$) optical depth of the simulated galaxies. 
The parameter $\delta$ is the fraction of the thickness of the central star+dust slab in the total thickness.
For the case of $\delta = 1$, Eq. \ref{eq:ep} corresponds to the well-mixed dusty slab geometry. 
On the other hand, it is the case with a central infinitely thin dusty sheet when $\delta$ is zero. 
We calculate the optical depth $\tau_{\rm d}$ using the following equation: 
\begin{equation}
\tau_{\rm d} = \frac{3 \Sigma_{\rm d}}{4a_{\rm d} s}, 
\end{equation} 
where $a_{\rm d}$ and $s$ are a typical size and the material density of dust grains, respectively. 
We set $a_{\rm d} = 0.1 ~{\rm \mu m}$ and $s = 2.5 ~{\rm g}~{\rm cm}^{-3}$ motivated by SNe dust production models \citep{Todini2001, Nozawa2003}. 
The dust surface density is defined by the following equation: 
\begin{equation}
\Sigma_{\rm d} = \frac{M_{\rm d}}{\pi r_{\rm d}^2}, 
\label{eq:dsmd}
\end{equation}
where $M_{\rm d}$ and $r_{\rm d}$ are the total dust mass and the effective radius of the dust distribution in a galaxy, respectively. 
These values are directly obtained from our simulations as follows. 
The dust mass $M_{\rm d}$ and $r_{\rm d}$ are assumed to be proportional to the metal mass, $M_{\rm metal}$, 
and the half stellar mass radius, $r_{\rm half}$, of each simulated galaxy, respectively. 
Therefore, $M_{\rm d}$ and $r_{\rm d}$ are expressed as $e_{\rm Md} M_{\rm metal}$ and $e_{\rm rd} r_{\rm half}$, 
where $e_{\rm Md}$ and $e_{\rm rd}$ are the proportional constants of the dust mass and the effective radius, respectively. 
These two parameters can be reduced to one parameter. 
Thus, equation~(\ref{eq:dsmd}) is reduced to 
\begin{equation}
\Sigma_{\rm d} = e_{\tau} \frac{M_{\rm metal}}{\pi r_{\rm half}^2}\,,
\end{equation}
where $e_{\tau} = e_{\rm Md} / e_{\rm rd}^2$ is a global constant for all the galaxies in our simulation. 
We calibrate $e_{\tau}$ and $\delta$ so as to reproduce the observed UV luminosity function (LF) at $z = 7$, and we keep these values even at higher redshifts. 
In this study, we adopt $e_{\tau} = 0.01$ and $\delta = 0.95$, respectively. 
When we calculate band magnitudes, 
we also apply the IGM absorption for the blue side of $1216 ~{\rm \AA}$ following \citet{Madau1995}, \citet{Inoue2014b} with an extrapolation for $z > 7$. 
The top panel of Fig. \ref{UVLF} represents the UV LFs from $z = 7$ to $z = 10$. 
Observational results are also shown in the same panel. 
Our model reproduces the observations well up to $z \sim 10$. 
Moreover, we predict the UV LFs beyond $z = 10$ in the bottom panel of Fig. \ref{UVLF}. 

Next, we consider the dust attenuation for various emission lines. 
In this study, we apply the same attenuation law as the continuum to the emission lines but with a different normalization \citep{Calzetti2000}. 
According to \citet{Calzetti2000}, the attenuation for emission lines is a factor of about 2 larger than that for the continuum at the same wavelength, although different factors were suggested by recent studies \citep[e.g.,][]{Capak2015}. 
The line escape probability ($f_{\rm UV}^{\rm line}$) at $1500~{\rm \AA}$ which is the normalization of the attenuation law for emission lines is 
\begin{equation}
f_{\rm UV}^{\rm line} = 10^{- 2.27\log{f_{\rm UV}^{\rm cont}}}, 
\label{eq:ep_line}
\end{equation}
where $f_{\rm UV}^{\rm cont}$ and 2.27 are the normalization of the attenuation law for the continuum (Eq. \ref{eq:ep}) 
and the attenuation enhancement for emission lines, respectively. 

\begin{figure}
\includegraphics[width = 80mm]{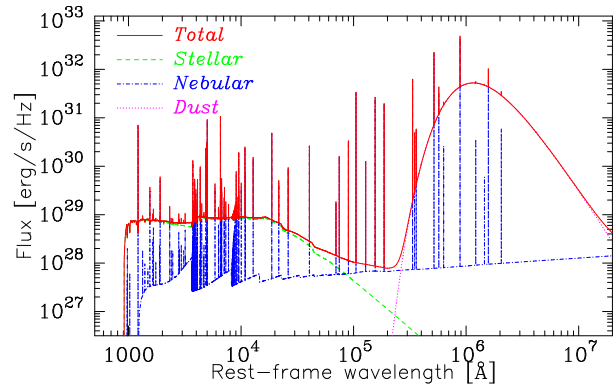}
\caption{The spectral energy distributions of a simulated galaxy without the IGM attenuation. 
The solid, dashed, dot-dashed and dotted lines represent the SED of total components (stellar + nebular + dust), stellar component, nebular components (continuum + line) and dust emission, respectively.} 
\label{SED}
\end{figure}

After calculation of the dust attenuation for the continuum and lines, 
we finally calculate the dust thermal emission in the same manner as \citet{Shimizu2012} and \citet{Shimizu2014}. 
One difference from our previous work \citep{Shimizu2012, Shimizu2014} is that not only the continuum photon energy but also the line photon energy absorbed by dust is converted into the dust luminosity. 
We assume that all the energy absorbed by the dust is re-emitted in the IR emission. 
The monochromatic luminosity ($L_{\nu}^{\rm dust}$) at frequency $\nu$ is written as 
\begin{equation}
L_{\nu}^{\rm dust} = 4\pi M_{\rm d} \kappa_{\nu} B_{\nu}(T_{\rm d}), 
\end{equation}
where $\kappa_{\nu}$, $B_{\nu}$ and $T_{\rm d}$ are the absorption coefficient, the Planck function and the dust grain temperature, respectively. 
We assume that 50 \% of the total metal mass in a simulated galaxy is converted into the dust mass ($M_{\rm d}$). 
The absorption coefficient $\kappa_{\nu}$ in the FIR is well described by a power-law $\kappa_{\nu} \propto \nu^{\beta}$ with $\beta = 1 \sim 2$. 
In this study, we apply $\beta = 1.7$ based on observational results \citep{Dunne2000, Dunne2001}. 
The typical dust temperature of our simulated galaxies is about $30~{\rm K}$. 
After these procedures, we obtain realistic SED of simulated galaxies from UV to IR/FIR wavelength. 
Fig. \ref{SED} represents an example SED from UV to IR of a simulated galaxy. 
As described our previous work \citep{Inoue2014a}, the [O {\sc iii}] $88 \rm \mu m$ line is the most prominent line. 
In the UV to optical range, many bright lines are also seen. 
We note that, in this study, the PDR/molecular cloud lines are not considered and the PAH features are not included in our SEDs, either. 
These are put into our future work. 

\subsection{Basic Physical Properties of Simulated Galaxies}
Here, we explore whether our simulation is reasonable to predict line detectability. 
We check essential physical properties of our simulated galaxies such as the stellar mass, the SFR, the dust attenuation and the metallicity which are necessary to calculate the line fluxes. 
In order to directly compare our model with the observations, 
we make galaxies distribution on a light-cone from $z = 6$ to $z = 17$ the same as in our previous work \citep{Shimizu2014}. 
The field-of-view (FOV) in this study is ($0.16~ \rm deg^2$) which is about 150 times wider than the HST FOV. 
We regard objects brighter than $H_{160} = 32$ as our selected galaxies. 
Then, we divide them into four redshift subsets at $z \sim 7~ (6.5 < z \leq 7.5), 8~ (7.5 < z \leq 8.5), 9~ (8.5 < z \leq 9.5)$ and $10~ (9.5 < z \leq 10.5)$.

\subsubsection{Star Formation Rate}
\begin{figure}
\includegraphics[width = 80mm]{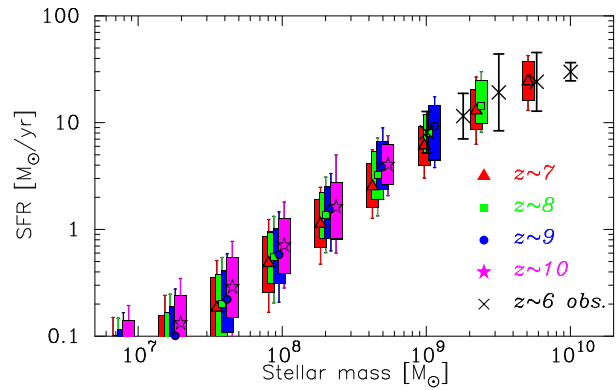}
\caption{Star formation rates as a function of the stellar mass. 
The large triangle, square, circle and star points are the median values of SFR in the stellar mass bins for $z \sim 7, ~8, ~9$ and $z \sim 10$,  receptively.  
The boxes and error-bars ranges show $68~\%$ percentile and $90~\%$ percentile in the distributions. 
The observational data at $z \sim 6$ also is shown as cross points with error-bars \citep{Salmon2015}. } 
\label{SM-SFR_HST}
\end{figure}

The Star formation rate (SFR) of simulated galaxies is very important 
because the line luminosities of various elements are essentially proportional to the SFR.
Fig. \ref{SM-SFR_HST} shows the star formation rates as a function of the stellar mass. 
The large triangle, square, circle and star points are the median values of SFR in the stellar mass bins for $z \sim 7, ~8, ~9$ and $z \sim 10$,  receptively.  
The boxes and error-bars ranges show $68~\%$ percentile and $90~\%$ percentile in the distributions. 
We also plot the observational data at $z \sim 6$ (cross points with error-bars) for comparison \citep{Salmon2015}. 
The SFR strongly correlates with their stellar mass. 
Interestingly, the trend is similar to that of the star forming galaxies at lower-$z$ Universe known as the main sequence distribution \citep[e.g.,][]{Daddi2007, Noeske2007}. 
We find that the distribution of $z \sim 7$ galaxies is almost same as the observation at $z \sim 6$. 
Moreover, the evolution from $z \sim 7$ to $z \sim 10$ is very weak. 
This implies that line luminosities are also proportional to their stellar mass. 

\subsubsection{Dust Attenuation}
\begin{figure}
\includegraphics[width = 80mm]{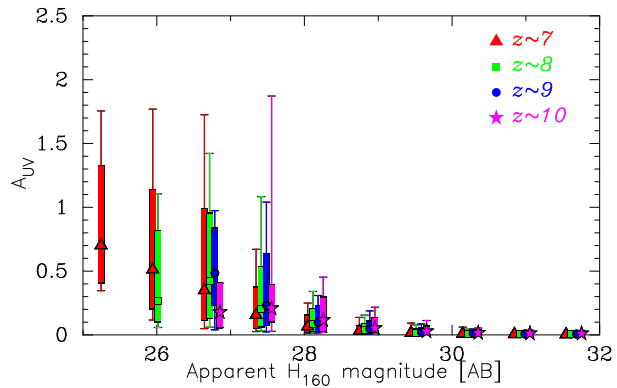}
\caption{The dust attenuation $A_{\rm UV}$ at $1500~{\rm \AA}$ as a function of $H_{160}$ band magnitude. 
The large triangle, square, circle and star points are the median values of $A_{\rm UV}$ in $H_{160}$ band magnitude bins for $z \sim 7, ~8, ~9$ and $z \sim 10$ , receptively. 
The boxes and error-bars ranges show $68~\%$ percentile and $90~\%$ percentile in the distributions. } 
\label{BAND-AUV_HST}
\end{figure}

Fig. \ref{BAND-AUV_HST} represents the dust attenuation $A_{\rm UV}$ at $1500~{\rm \AA}$ as a function of $H_{160}$ band magnitude. 
The large triangle, square, circle and star points are the median values of $A_{\rm UV}$ in $H_{160}$ band magnitude bins for $z \sim 7, ~8, ~9$ and $z \sim 10$ , receptively. 
The boxes and error-bars ranges show $68~\%$ percentile and $90~\%$ percentile in the distributions. 
Clearly, there are a few galaxies strongly affected by the dust attenuation. 
The typical value of the dust attenuation correlates with the UV magnitude even though there is a large scatter.  
We note that the typical value is smaller than that we find in our previous work \citep{Shimizu2014}. 
In the new code which we use in this study, the radiation pressure and the AGN feedback are newly included in addition to the SN feedback \citep{Okamoto2014}. 
Especially, the radiation pressure and the SN feedback effectively suppress the star formation in small mass galaxies than the case of the SN feedback only simulation. 
As a result, less dust attenuation is required to reproduce the UV LFs. 

\begin{figure*}
\includegraphics[width = 160mm]{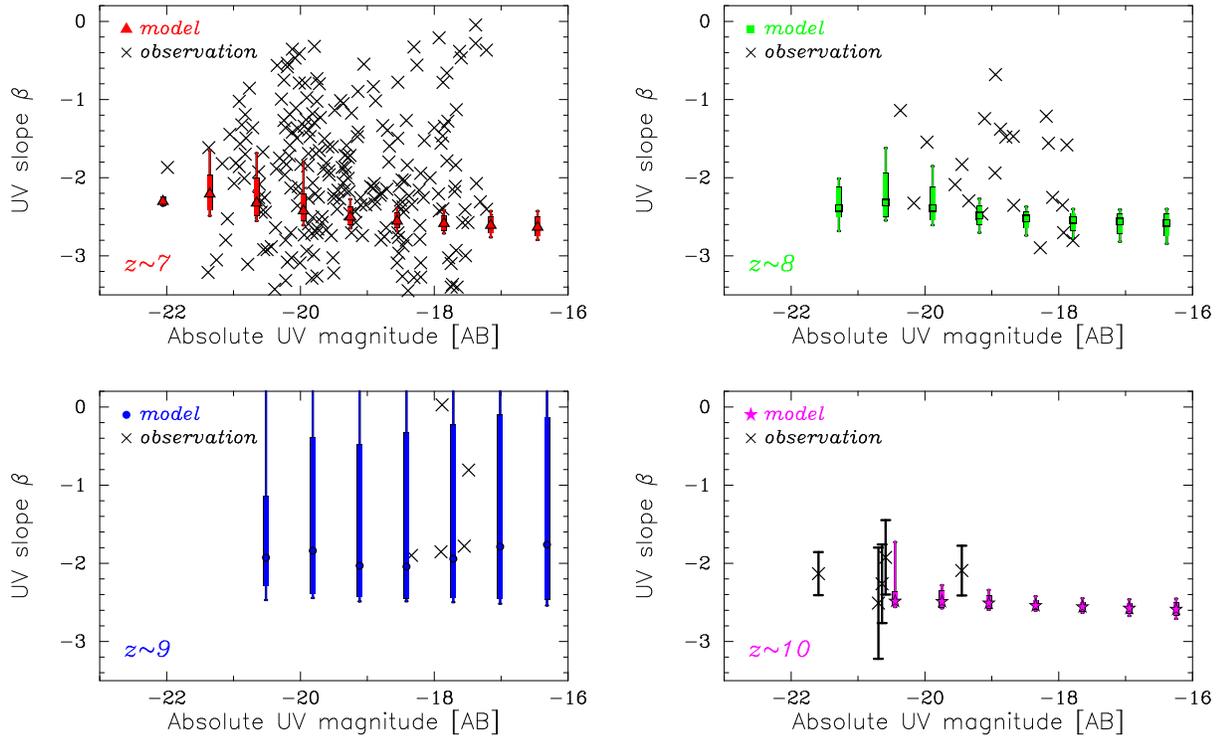}
\caption{The UV slope $\beta$ as a function of the absolute UV magnitude. 
The large triangle, square, circle and star points are the median values of $\beta$ in the absolute UV magnitude bins for $z \sim 7, ~8, ~9$ and $z \sim 10$, receptively.  
The boxes and error-bars ranges show $68~\%$ percentile and $90~\%$ percentile in the distributions. 
The large cross points are shown the observational results \citep{Dunlop2013, Bouwens2014b, Wilkins2016}} 
\label{UV_SLOPE_HST}
\end{figure*}

Next, we calculate the UV slope $\beta$ with the exactly same formulae as that in the observations \citep{Dunlop2013, Bouwens2014b, Wilkins2016}: 
\begin{equation}
\beta = 4.39 \times (J_{125} - H_{160})- 2.0, 
\end{equation}
for $z \sim 7$ galaxies, 
\begin{equation}
\beta = 8.98 \times (JH_{140} - H_{160})- 2.0, 
\end{equation}
for $z \sim 8$ galaxies, 
\begin{equation}
\beta = 9.32 \times (JH_{140} - H_{160})- 2.0, 
\end{equation}
for $z \sim 9$ galaxies and 
\begin{equation}
\beta = 1.1 \times (H_{160} - [3.6])- 2.0, 
\end{equation}
for $z \sim 10$ galaxies, respectively. 
$J_{125}$ and $JH_{140}$ are WFC3 band filters equipped on the Hubble Space Telescope and $[3.6]$ is the Spitzer IRAC band1. 
Fig. \ref{UV_SLOPE_HST} shows the slope $\beta$ as a function of the absolute UV magnitude (without dust correction). 
The large triangle, square, circle and star points are the median values of $\beta$ in the absolute UV magnitude bins for $z \sim 7, ~8, ~9$ and $z \sim 10$, receptively.  
The boxes and error-bars ranges show $68~\%$ percentile and $90~\%$ percentile in the distributions. 
The large cross points are also shown as the observational data \citep{Dunlop2013, Bouwens2014b, Wilkins2016}. 
As described our previous paper, the large scatter of the observational data may be due to the observational error and/or uncertainty. 
The UV brighter galaxies have larger $\beta$ values. 
This is because the UV brighter galaxies tend to be more evolved and have a smaller fraction of young stellar populations than the UV fainter ones. 
Moreover, such UV brighter galaxies rapidly proceed metal enrichment. 
This trend can be seen in high-$z$ galaxy observations \citep{Bouwens2014b, Wilkins2016}. 
At $z \sim 9$, the $\beta$ distribution of our model shows the large scatter and different trend from the other redshift ones. 
This is because Ly$\alpha$ break or line of some $z \sim 9~(8.5 < z < 9.5)$ simulated galaxies enters in $JH_{140}$ band. 
If we adopted the same $\beta$ formula as $z \sim 10$ for $z \sim 9$, we would get a similar $\beta$ distribution for $z \sim 9$ to other redshifts. 
This suggests that some modifications are necessary for $z \sim 9$ $\beta$ estimator. 

\subsection{Metallicity}
\begin{figure}
\includegraphics[width = 80mm]{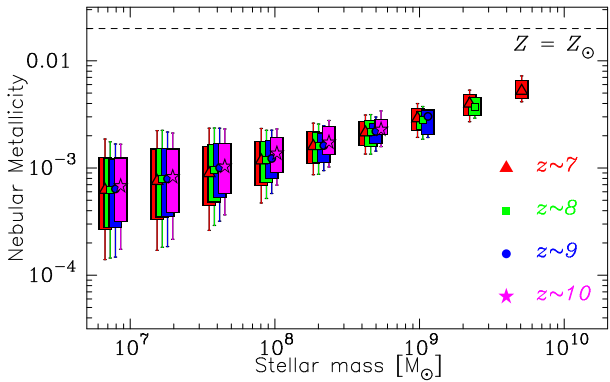}
\caption{The nebular metallicity as a function of the stellar mass.
The large triangle, square, circle and star points are the median values of the metallicity  in the stellar mass bins for $z \sim 7, ~8, ~9$ and $z \sim 10$, receptively.  
The boxes and error-bars ranges show $68~\%$ percentile and $90~\%$ percentile in the distributions. 
The dashed line is solar metallicity $Z_{\odot} = 0.02$. }
\label{SM-METAL_HST}
\end{figure}

In the fiducial line emission model of \citet{Inoue2011}, 
the line emissivities are proportional to the metallicity in the case of $Z < 0.1~Z_{\odot}$ where $Z$ and $Z_{\odot}$ are the gas metallicity and the solar metallicity ($Z_{\odot} = 0.02$).
Then, when $Z > 0.1~Z_{\odot}$, the dependence becomes opposite because of lower Lyman continuum (LyC) emissivity for higher metallicity.
Thus, in order to detect the line emissions of very high-$z$ galaxies, 
the follow-up survey for $\sim 0.1 Z_{\odot}$ systems is efficient. 
Fig. \ref{SM-METAL_HST} represents the nebular metallicity of the simulated galaxies as a function of the stellar mass. 
The definition of the metallicity is the same as our previous work \citep{Shimizu2014} in which the metallicity is defined by the LyC luminosity weighted average metallicity called the nebular metallicity rather than the mass weighted average metallicity.
The nebular metallicity of the simulated galaxies is proportional to their stellar mass.
The metallicity of some galaxies reaches $\sim 0.1 Z_{\odot}$ even at $z \sim 10$. 
This means that the metal enrichment or the chemical evolution proceeds rapidly in the early Universe. 
Interestingly, the redshift dependence of the metallicity is very weak or almost no evolution. 
Thus, the each line emissivity at the redshift range varies little if their stellar mass is similar values. 

\section{Results and Discussion}
In this section, using our mock galaxies, we explore expected fluxes of some emission lines. 
We investigate the observed line fluxes of our simulated galaxies and discuss the detectability of these lines at $z = 7$--10. 
Then, we predict the detectability in $z > 10$ galaxies.

\subsection{Line Flux Prediction}
\begin{figure*}
\includegraphics[width = 160mm]{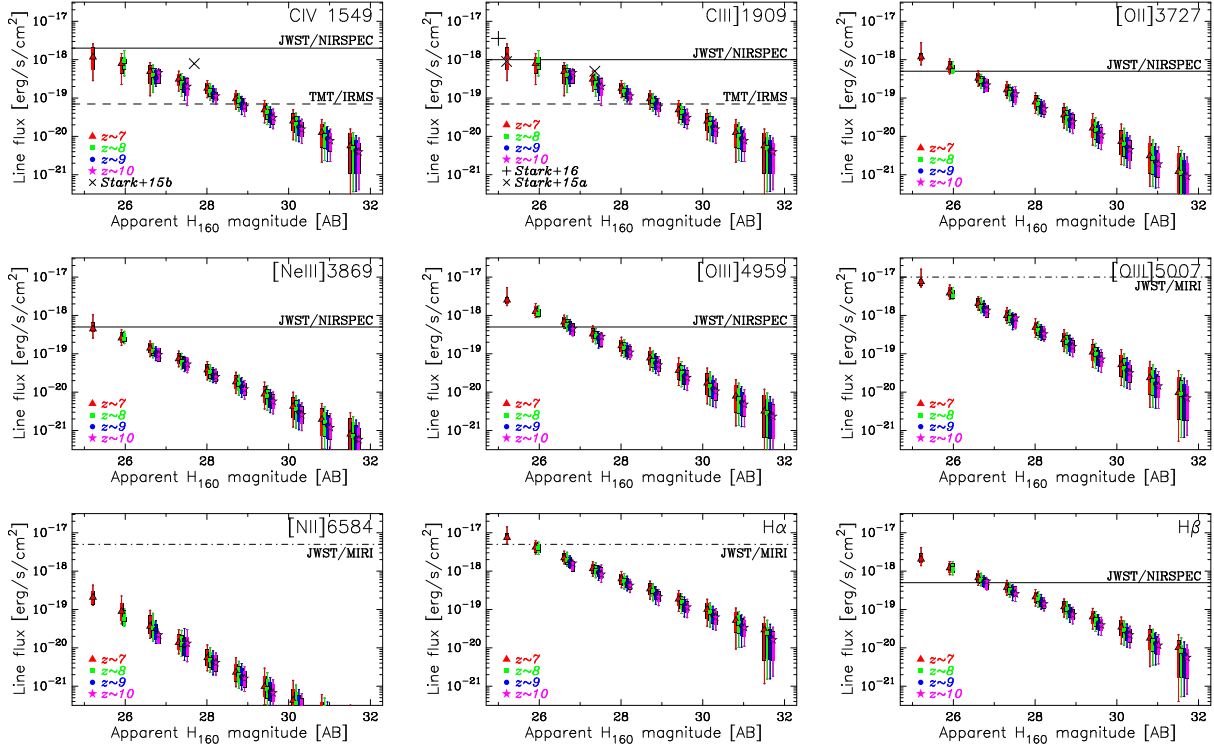}
\caption{The observed line flux in the rest-frame UV to optical wavelength as a function of the apparent $H_{160}$ band magnitude. 
The large triangle, square, circle and star points are the median values of line fluxes in the $H_{160}$ band magnitude bins for $z \sim 7, ~8, ~9$ and $z \sim 10$, receptively. 
The boxes and error-bars ranges show $68~\%$ percentile and $90~\%$ percentile in the distributions. 
The large cross and plus-sign points are observational results \citep{Stark2015a, Stark2015b, Stark2016}. 
The solid, dot-dashed and dashed lines are the detection limits of the JWST/NIRSPEC ($\lambda_{\rm obs}= 0.6$--5 $[\rm \mu m]$), the JWST/MIRI ($\lambda_{\rm obs}= 5$--28 $[\rm \mu m]$) and the TMT/IRMS ($\lambda_{\rm obs}= 0.8$--2.5 $[\rm \mu m]$) at the wavelength of the line for a $z = 9$ galaxy, respectively, 
if the integration time is $10^4~\rm [sec]$ and the signal-to-noise ratio (S/N) is 5. } 
\label{BAND-LINE_HST}
\end{figure*}

Here, we show some bright and important metal lines for the diagnostics of the gas metallicity. 
We also present hydrogen recombination lines such as H$\alpha$ and H$\beta$. 
The $R23$-index defined by ([O {\sc iii}] ${4959 \rm \AA}$+[O {\sc iii}] ${5007 \rm \AA}$+[O {\sc ii}] ${3727 \rm \AA}$)/H$\beta$ is well known as a good indicator for gas metallicity. 
Moreover, the N2 ($\log$([N {\sc ii}] $6584 \rm \AA$/H$\alpha$)) and O3N2 ($\log$(([O {\sc iii}] $5007 \rm \AA$/H$\beta$)/([N {\sc ii}] $6584 \rm \AA$/H$\alpha$))) indices are also used to estimate the oxygen abundances. 
\citet{Nagao2006} suggests that the line ratio [Ne {\sc iii}] $3869 \rm \AA$/[O {\sc ii}] $3727 \rm \AA$ is also a useful metallicity indicator for high-$z$ galaxies, 
especially when the R23-index or other diagnostics involving [O {\sc iii}] $5007 \rm \AA$ or [N {\sc ii}] $6584 \rm \AA$ are not available. 
Fig. \ref{BAND-LINE_HST} represents the observed line fluxes of those important emission lines in the rest-frame UV to optical wavelength as a function of observed $H_{160}$ band magnitude. 
We calculate all elements for the diagnostics (R23-index, N2 and O3N2) of the gas metallicity. 
The small triangle, square, circle and star points are the expected line fluxes of the simulated galaxies at $z \sim 7$, 8, 9 and 10, respectively, 
while the large points with error-bars represent the average values and their standard deviations in $H_{160}$ band magnitude bins. 
We also plot the observational results by \citet{Stark2015a}, \citet{Stark2015b} and \citet{Stark2016} as the two cross and plus-sign points (the fainter point was demagnified by the quoted lensing magnification) in the C{\sc iv} $1549 \rm \AA$ and the C {\sc iii}] ${1909 \rm \AA}$ panels, which are very consistent with our prediction. 

We explore whether the existing facilities such as the VLT/X-Shooter and the Keck/MOSFIRE can detect the emission lines considered in this paper. 
The wavelength range of these instruments is about 1--2.5 $\rm \mu m$. 
Although, the detection limit is not deep enough (a few$\times 10^{-18}~\rm [erg/s/cm^2]$) under a practical observational condition, 
the C{\sc iv} $1549 \rm \AA$ and the C {\sc iii}] ${1909 \rm \AA}$ lines of the brighter galaxies ($< 26~\rm mag$) are detectable at $z > 7$ without a significant magnification by gravitational lensing. 
Recently, \citet{Zitrin2015a} reported no C {\sc iii}] ${1909 \rm \AA}$ line with $H_{160} \sim 28$ was detected under the condition of the detection limit $1.5 \times 10^{-18}~\rm [erg/s/cm^2]$. 
This is also consistent with our result because typical line fluxes with $H_{160} \sim 28$ are $2 \times 10^{-19}~\rm [erg/s/cm^2]$ which is below the observational detection limit. 

In order to detect these nebular lines, we need the future facilities with a wider wavelength coverage (the rest-frame UV to optical) and a stronger light-gathering power. 
The JWST and the TMT have potential to detect the nebular line of very high-$z$ galaxies. 
We consider the detectability of the nebular lines using the JWST and the TMT. 
The JWST/NIRSPEC, the JWST/MIRI and the TMT/IRMS provide spectroscopy over the wavelength range of 0.6--5 $\rm \mu m$, 5--28 $\rm \mu m$ and 0.8--2.5 $\rm \mu m$, respectively. 
The point-source detection limits of the JWST/NIRSPEC, JWST/MIRI and TMT/IRMS, which depend on the observed wavelength are a few$\times 10^{-19}~ \rm [erg/s/cm^2]$, a few$\times 10^{-18}~\rm [erg/s/cm^2]$ and a few$\times 10^{-20}~ \rm [erg/s/cm^2]$, respectively, 
if the integration time is $1.0 \times 10^4~ \rm [sec]$ and the signal to noise ratio (S/N) is 5. 
We note that only C {\sc iv} ${1549 \rm \AA}$ and the C {\sc iii}] ${1909 \rm \AA}$ lines at $z = 7$--10 can be available for the TMT/IRMS due to the wavelength coverage shorter than 2.5 $\rm \mu m$. 
Interestingly, the redshift evolution of the each line flux against the apparent $H_{160}$ band magnitude is not seen. 
This is because the SFR and metallicity evolutions of the simulated galaxies are weak 
as shown in Fig. \ref{SM-SFR_HST} and Fig. \ref{SM-METAL_HST}. 
Moreover, the evolution of the dust attenuation is also weak as shown in Fig. \ref{BAND-AUV_HST}. 
Almost all these lines are detectable by these instruments if the band magnitude are brighter than about 28 mag. 
This means that we may detect some lines even at $z > 10$ galaxies if the galaxies are bright enough ($< 28 ~\rm mag$) 
and we can confirm the spectroscopic redshift using these lines. 
Moreover, we can study the gas metallicity of bright high-$z$ galaxies ($< 28~ \rm mag$) through the line ratio diagnostics. 
Finally, we comment on He{\sc ii} ${1640 \rm \AA}$ in the CR7 \citep{Sobral2015b}. 
We also calculate the line and search the CR7 candidates around $z = 6.6$. 
However, we can not find any CR7 like He{\sc ii} ${1640 \rm \AA}$ bright galaxies in our simulation. 
This suggests that in order to reproduce such an object, we should consider the first star formation and/or intermediate black-hole formation \citep[e.g.,][]{Agarwal2015, Hartwig2015, Pacucci2015, Pallottini2015}. 
This is beyond the scope of this paper. It will be our future work. 

\subsection{Line Equivalent Width Prediction}
\begin{figure*}
\includegraphics[width = 160mm]{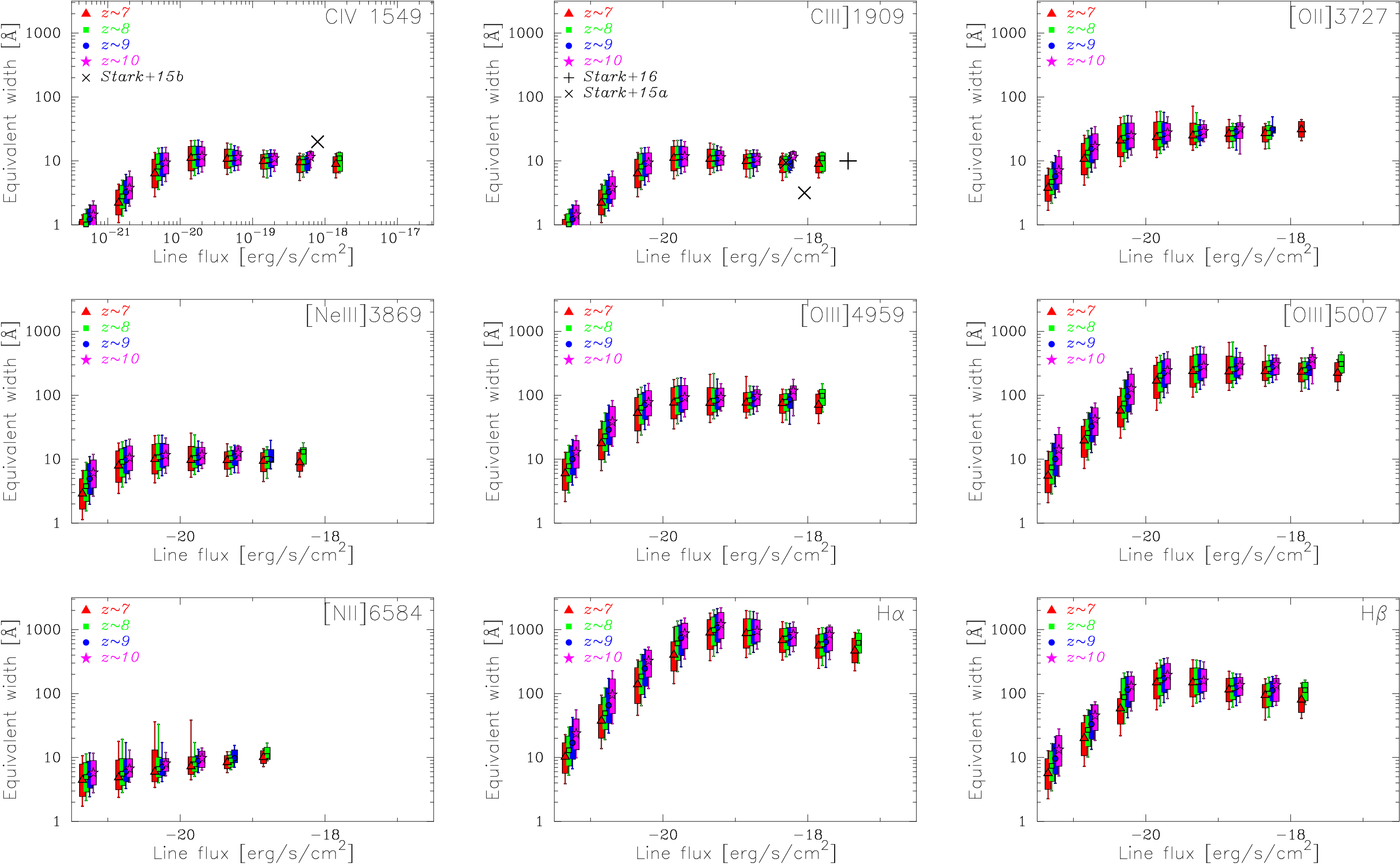}
\caption{The rest-frame EW of various lines in the rest-frame UV to optical wavelength as a function of observed line fluxes. 
The large triangle, square, circle and star points are the median values of EWs in the observed line flux bins for $z \sim 7, ~8, ~9$ and $z \sim 10$, receptively.  
The boxes and error-bars ranges show $68~\%$ percentile and $90~\%$ percentile in the distributions. 
The large cross and plus-sign points are observational results \citep{Stark2015a, Stark2015b, Stark2016}. } 
\label{FLUX-EW_HST}
\end{figure*}

The equivalent width (EW) of lines is also very interesting to discuss detectability of lines in high-$z$ galaxies because the value strongly depends on the star formation history and the IMF. 
Fig. \ref{FLUX-EW_HST} represents the rest-frame EWs of various lines in the rest-frame UV to optical wavelength as a function of line fluxes. 
The large triangle, square, circle and star points are the median values of EWs in the observed line flux bin for $z \sim 7, ~8, ~9$ and $z \sim 10$, receptively. 
The boxes and error-bars ranges show $68~\%$ percentile and $90~\%$ percentile in the distributions. 
We also plot the observational results by \citet{Stark2015a}, \citet{Stark2015b} and \citet{Stark2016} as the cross and plus-sign points in the C{\sc iv} $1549 \rm \AA$ and the C {\sc iii}] ${1909 \rm \AA}$ panels. 
The strong lines exceeding $100~\rm\AA$ can be seen in some lines such as [O{\sc iii}] $4959/5007 \rm \AA$, H$\alpha$ and H$\beta$. 
In the fainter line fluxes (less massive galaxies) regime ($< \sim 10^{20}~ \rm [erg/s/cm^2]$), the EWs are proportional to their line fluxes. 
The EW represents the current SFR divided by an average of SFR over a certain time duration, and the line flux is almost proportional to the SFR. 
The metallicity effect on the line emissivity is secondary because the simulated galaxies have more or less a metallicity about $\sim 0.1$ solar (see Fig. \ref{SM-METAL_HST}). If we choose galaxies emitting a weak line flux, they are in a low SFR phase, so that the EW is small.
In the brighter line fluxes (massive galaxies) regime ($> \sim 10^{20}~ \rm [erg/s/cm^2]$), 
EWs are almost constant (or very slowly decrease with increasing line fluxes). 
These galaxies are in an on-going star-formation episode (or starburst) and their EW (or $\rm SFR/<SFR>$) reaches an asymptotic value.
Interestingly, only the H$\alpha$ and H$\beta$ lines show a decreasing EW as increasing the line flux rather than a constant. 
This is because stronger line flux galaxies tend to have a higher metallicity and the LyC photons production efficiency per nonionizing photon decreases with the metallicty increasing. 
On the other hand, for metal lines, the effect is cancelled out by an increase of the line emissivity due to increasing metallicity. 
Finally, we mention higher H$\alpha$ EW galaxies than theoretical expectation reported by \citet{Shim2011} and \citet{Smit2016}. 
In our simulation, there is no object which can reproduce such a very high EW. 
This implies that in order to reproduce such higher EW galaxies, we may need to include the metal-free star formation and/or a top-heavy IMF which we are not considered in this study. 

\subsection{Detectability of the Lines beyond $z\sim10$}
We have presented the detectability of multiple emission lines in the rest-frame UV to optical wavelength at  $z \sim 7$--10. 
Here, we discuss the detectability of these lines at even higher-$z$, $z = 11$--15. 
The number of bright galaxies decreases with increasing redshift, 
thus, a very wide survey is necessary to discover sufficiently bright very high-$z$ galaxies. 
In this study, we feature high-$z$ galaxies which will be detected with 
the Wide-Field Infrared Survey Telescope (WFIRST)\footnote{http://wfirst.gsfc.nasa.gov/} 
and the First Light And Reionization Explorer (FLARE)\footnote{http://mission.lam.fr/flare/}. 
The surveys with these facilities are potentially the most efficient one to detect very high-$z$ galaxies thanks to the wide survey area and the depth.
We notice that the instrument design of these facilities has not been determined yet. 
In the survey strategy, they expect to detect many high-$z$ galaxy candidates at $11 \leq z \leq 15$. 
In order to compare our model with the future observations, we make a mock galaxy catalogue at $11 \leq z \leq 15$. 
We can identify a few galaxies at $z \sim 15$ even though the FOV of our simulation is very narrow ($\sim 0.15~\rm deg^2$). 

\begin{figure}
\includegraphics[width = 80mm]{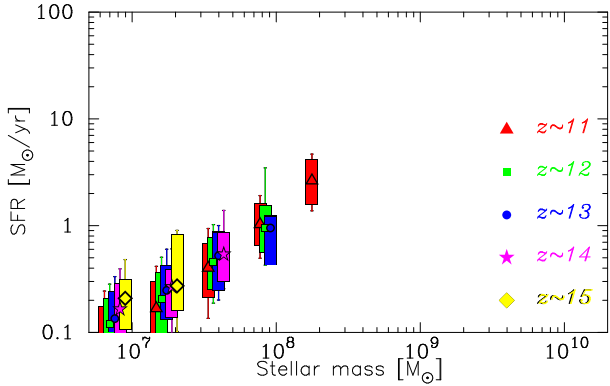}
\caption{Same as Fig. \ref{SM-SFR_HST}, but for $11 < z < 15$. }
\label{SM-SFR_HIGH_Z}
\end{figure}

\begin{figure}
\includegraphics[width = 80mm]{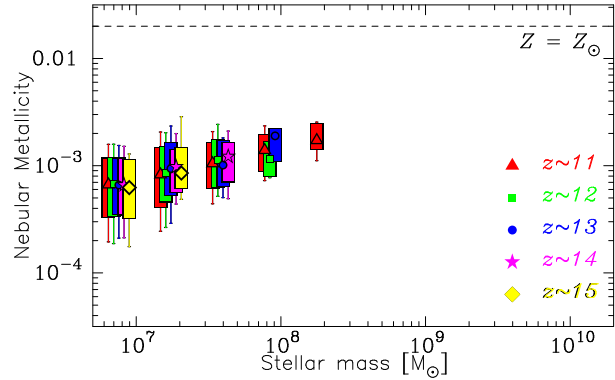}
\caption{Same as Fig. \ref{SM-METAL_HST}, but for $11 < z < 15$. }
\label{SM-METAL_HIGH_Z}
\end{figure}

Fig. \ref{SM-SFR_HIGH_Z} and \ref{SM-METAL_HIGH_Z} represent the evolution of SFR and nebular metallicity, respectively. 
These results suggest that a weak redshift evolution of the line flux--magnitude relation even at $z > 11$ as lower-$z$ discussed in the previous section. 
Thus, some line emissions with K band magnitude $ < 26$ mag can be detected even at $z \sim 15$ with the JWST/NIRSPEC and the JWST/MIRI. 
Next we estimate the line fluxes of these selected galaxies. 

\begin{figure*}
\includegraphics[width = 160mm]{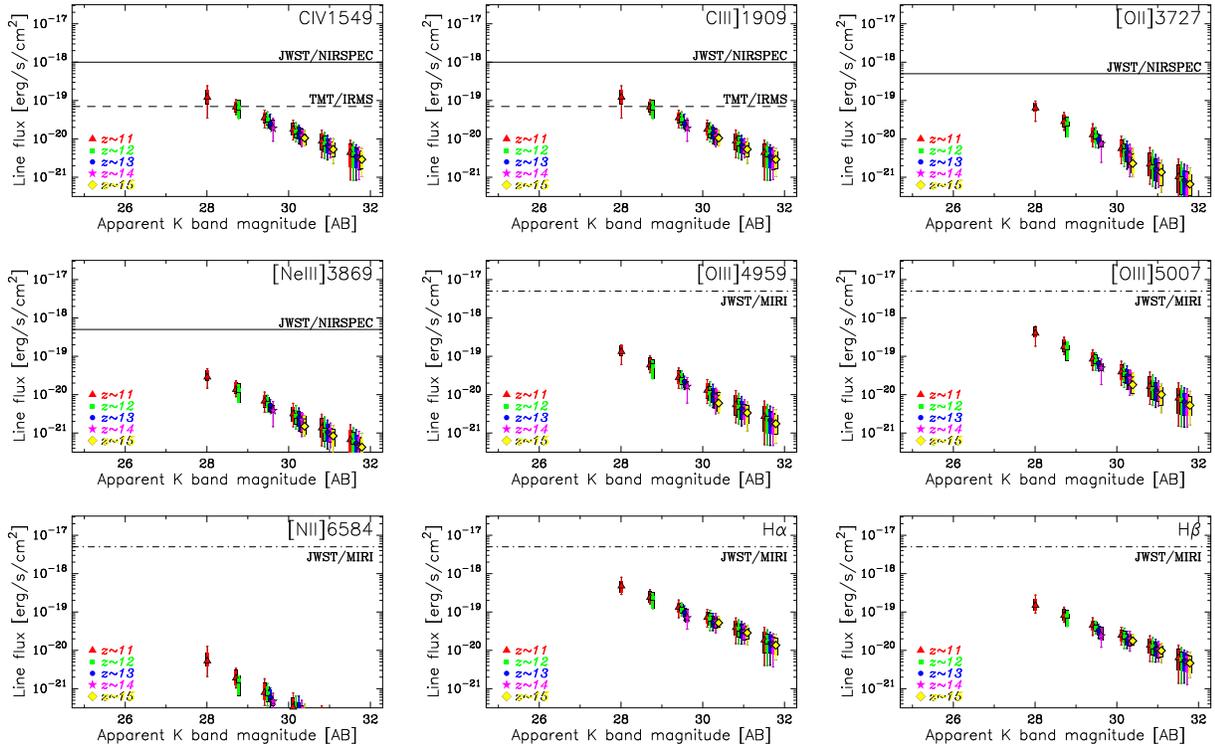}
\caption{Same as Fig. \ref{BAND-LINE_HST}, but for $11 < z < 15$. 
The solid, dot-dashed and dashed lines are the detection limit of the JWST/NIRSPEC, the JWST/MIRI and the TMT/IRMS at the wavelength of the line for a $z = 12$ galaxy, respectively, 
if the integration time is $10^4~\rm [sec]$ and the signal-to-noise ratio (S/N) is 5. } 
\label{BAND-LINE_HIGH_Z}
\end{figure*}

\begin{figure*}
\includegraphics[width = 160mm]{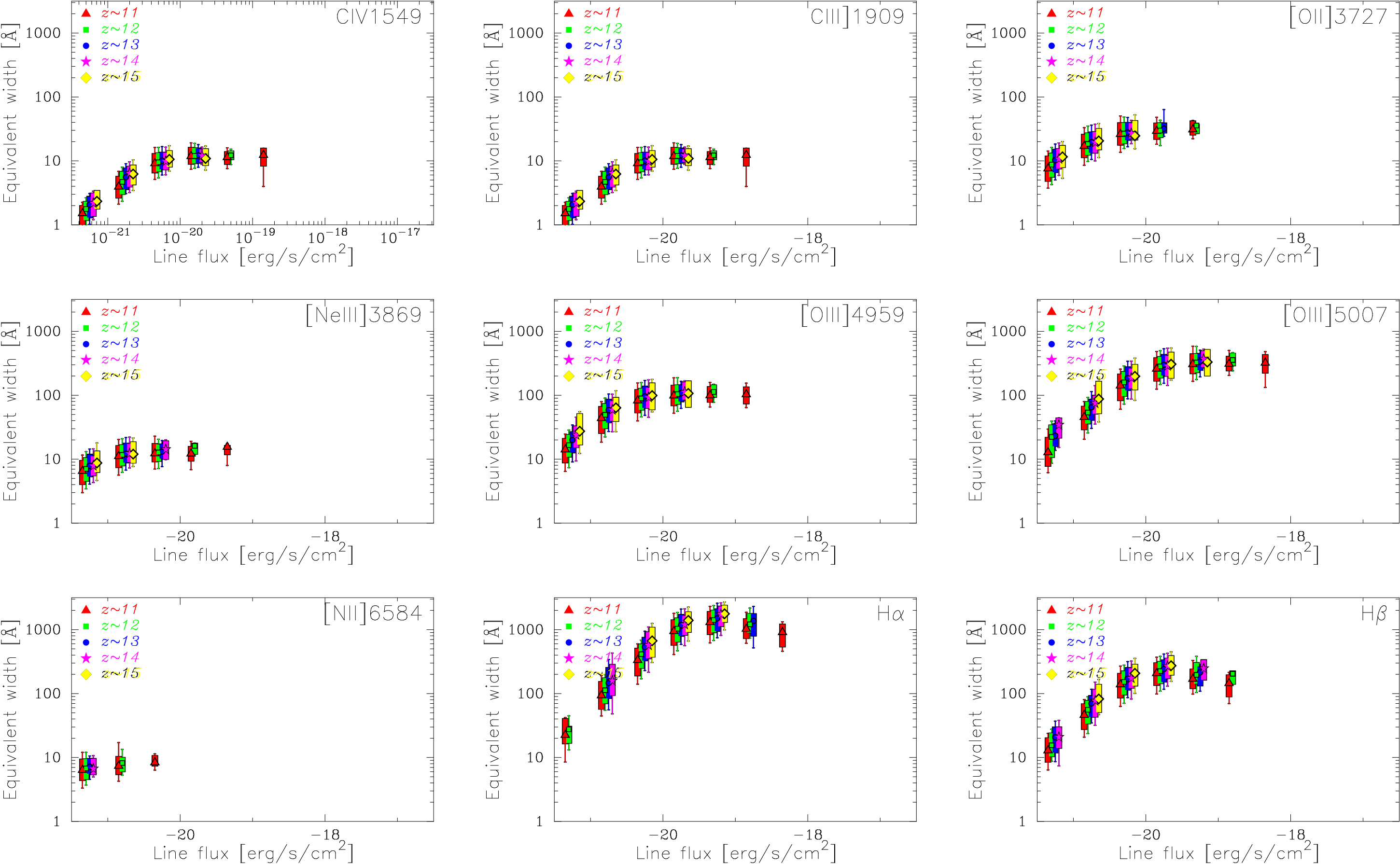}
\caption{Same as Fig. \ref{FLUX-EW_HST}, but for $11 < z < 15$. } 
\label{FLUX-EW_HIGH_Z}
\end{figure*}

Fig. \ref{BAND-LINE_HIGH_Z} represents the expected line fluxes in the rest-frame UV to optical wavelength as a function of K band magnitude. 
The point styles and their corresponding redshifts are noted in the panel. 
The points with error-bars represent the average values and the standard deviations in K band magnitude bins. 
The line flux--magnitude relations of different redshifts are almost the same as found in the cases of $z \leq 10$ galaxies as expected. 
We find that the C {\sc iv} $1549 \rm \AA$ and the C {\sc iii}] $1909 \rm \AA$ lines of simulated galaxies with K band magnitude $< 28$ mag are detectable even at $z \sim 12$ with the TMT/IRMS. 
On the other hand, in our simulation, there are no galaxies at $z \sim 15$ that are bright enough. 
This implies that the survey area of our simulation (FOV$\sim 0.15~\rm deg^2$) is not wide enough to have galaxies emitting line fluxes bright enough to be detectable with the JWST or the TMT. 
Therefore, a much wider survey area is imperative to discover line detectable galaxies beyond $z \sim 11$. 
We can detect some emission lines using the JWST or the TMT if the bright galaxy candidates are discovered at any redshifts because there is no evolution in the relation between apparent UV magnitude and line fluxes along the redshift. 
We also present the rest-frame EWs as a function of observed line fluxes in Fig. \ref{FLUX-EW_HIGH_Z}. 
Same trend as the $z < 10$ cases can be seen. 
There are very EW strong lines exceeding 100 $\rm \AA$ even $z > 10$. 

\subsection{Nebular Emission Line Luminosity Functions}
\begin{figure*}
\includegraphics[width = 160mm]{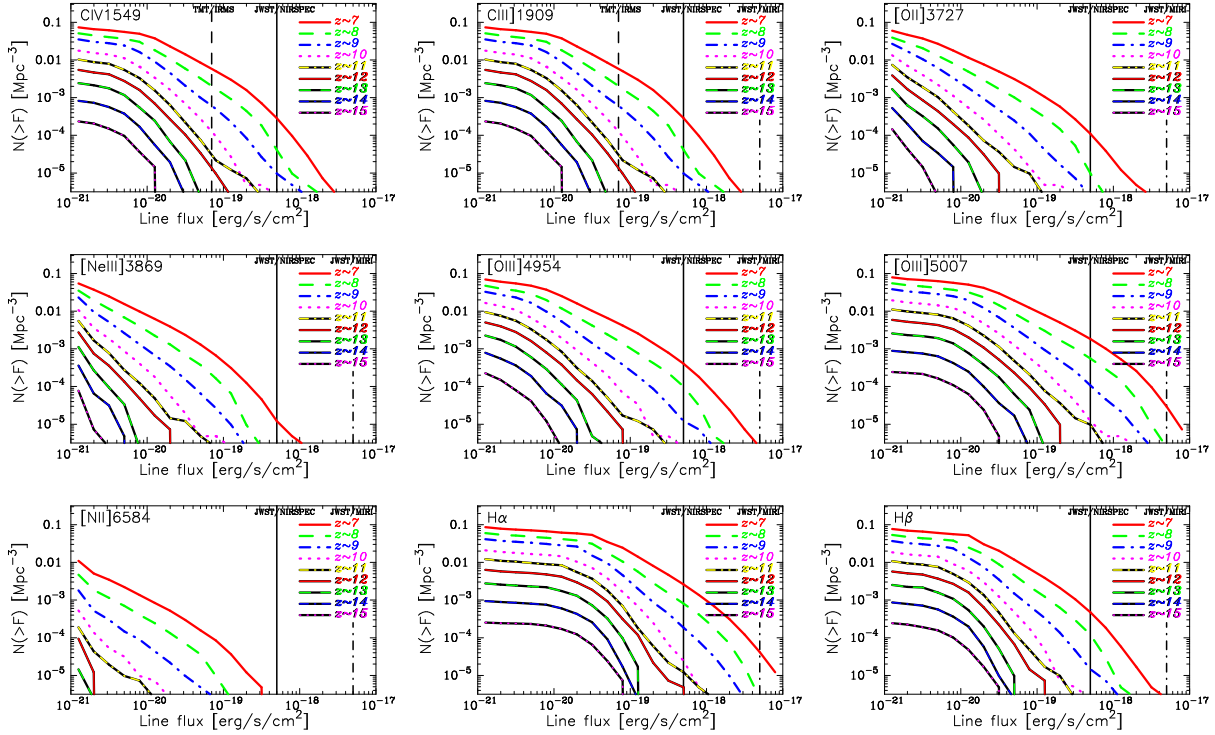}
\caption{The cumulative line luminosity functions at $7 < z < 15$. 
The line styles and their corresponding redshifts are noted in the panel. 
Vertical lines represent typical detection limit of TMT/IRMS (dashed line), JWST/NIRSPEC (solid line) and JWST/MIRI (dot-dashed line) assuming the integration time is $10^4~\rm [sec]$ and the signal-to-noise ratio (S/N) is 5, respectively. } 
\label{LINE_LF}
\end{figure*}

We have not discussed yet how many line emitting galaxies can be detected by the future telescopes. 
This topic is very important for future plans. 
We show cumulative luminosity functions (LFs) of various emission lines at $7 \leq z \leq 15$ in Fig. \ref{LINE_LF}. 
Here, we discuss only the C {\sc iv} $1549 \rm \AA$ and the C {\sc iii}] $1909 \rm \AA$ lines because the line have strong potential to be detected even at $z > 10$ with the TMT and the JWST.
Typical detection limits of TMT/IRMS (dashed line), JWST/NIRSPEC (solid line) and JWST/MIRI are also shown in Fig. \ref{LINE_LF} 
if the integration time is $10^4~\rm [sec]$ and the signal-to-noise ratio (S/N) is 5. 
The detection limit for TMT/IRMS is around $10^{-19}~\rm [erg/s/cm^2]$. 
Thus, one or more C {\sc iv} $1549 \rm \AA$ and C {\sc iii}] $1909 \rm \AA$ emitters at $z < 12$ can be detected by TMT/IRMS if the survey volume is about $10^5~\rm [Mpc^3]$ which is equivalent to $> 200$ FOVs of the TMT and the JWST. 
On the other hand, the wavelength of the line at $z > 12$ is beyond the coverage of the instrument. 
The JWST/NIRSPEC can cover that wavelength range and the detection limit is around $10^{-18}~\rm [erg/s/cm^2]$. 
In this case, a larger survey volume is necessary to detect the emitters because there is no detectable galaxy in our small FOV ($\sim 0.15~\rm deg^2$) simulation. 
We emphasise again that a wide imaging survey with the WFIRST and the FLARE is very useful to discover very higher-$z$ line emitters. 

\subsection{Gravitationally-lensed $z > 9$ Galaxy Candidates Found to Date}
\label{Discussion3}
Here, we mention detectability of galaxies magnified by gravitational lensing. 
Recently, many surveys with lensing magnification have been actively performed. 
Many faint galaxies have been discovered by the combination the HST/WFC3 and gravitational lensing. 
Thanks to the magnification by gravitational-lensing, some lensed galaxies which are candidates of $z > 9$ objects are detectable by even the current facilities such as the VLT/X-Shooter and the Keck/MOSFIRE. 

\begin{figure}
\includegraphics[width=80mm]{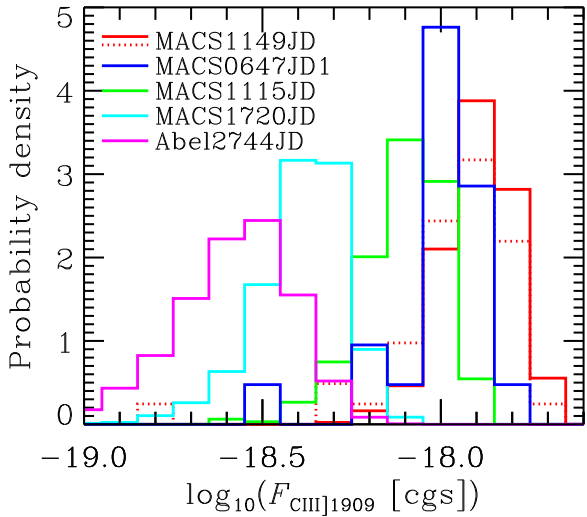}
\caption{Probability distribution of the expected flux of the C {\sc iii}] $1909 \rm \AA$ line for gravitationally-lensed $z > 9$ galaxy candidates found so far. For MACS1149JD, there are two estimates for the magnification factor and we show two cases (solid and dotted histograms).}
\label{LENSED_GALAXY}
\end{figure}

Let us discuss the detectability of the C {\sc iii}] $1909 \rm \AA$ emission line from $z > 9$ galaxy candidates found by surveys for galaxy cluster lensing fields like the Cluster Lensing And Supernova survey with Hubble (CLASH) and the Hubble Frontier Field (HFF) with instruments currently available.
The C {\sc iii}] $1909 \rm \AA$ line is the second strongest emission line after Ly$\alpha$ in the UV range and can be observed with ground-based 8-10 m class telescopes at $z < 12$.
Table \ref{TABLE1} is a summary of the candidates to be discussed here.
First, we estimate the range of their real apparent $H_{160}$ magnitudes by demagnifying with the quoted magnification factor $\mu$ and its uncertainty.
Next, we select the galaxies within the magnitude range and satisfying appropriate colour selection criteria from our simulation.
Then, we obtain the magnification factor for each simulated galaxy so as to be observed as the observed $H_{160}$ magnitude.
Finally, we magnify the C {\sc iii}] $1909 \rm \AA$ emission line flux by this factor, and obtain the distribution of the expected C {\sc iii}] $1909 \rm \AA$ line fluxes.
Fig. \ref{LENSED_GALAXY} shows the result for the five $z > 9$ candidates.
The currently reachable limit is $1\times10^{-18}~\rm [erg/s/cm^2]$ \citep[e.g.,][]{Stark2014}.
The C {\sc iii}] $1909 \rm \AA$ line from MACS1149JD is detectable with a high probability, 
73 or 85\% for the two quoted magnification factors.
The detectability of the line from MACS0647JD1 is also relatively high (57\%).
The other three candidates have a detectability of $ < 5$\%.
Therefore, MACS1149JD and MACS0647JD1 are the two best targets for the follow-up spectroscopy aiming to detect the C {\sc iii}] $1909 \rm \AA$ line even with the current facilities. 
In order to study higher-$z$ ($z > 7$) line emitting galaxies, 
the combination of the GMT, the TMT or the JWST and gravitational lensing effect could be one of the most powerful ways. 

\begin{table*}
\begin{tabular}{lcccccl}
\hline
Object & $z_{\rm ph}$ & $H_{160}$ [AB] & $\mu$ & $\log10{(F_{{\rm C III]}1909}/[\rm cgs])}$ & $P(< -18.0)$ [\%] & References \\
\hline
MACS1149JD & $9.6\pm0.2$ & $25.7\pm0.1$ & $14.5^{+4.2}_{-1.0}$ / $5.5^{+0.31}_{-0.28}$ & $-17.9 \pm 0.1$ & $85/73$ & Z12, B14, H15 \\
MACS0647JD1 & 10.20--11.47 & $25.88\pm0.09$ & $\sim8.4$ & $-18.0 \pm 0.1$ & $57$ & C13, GL14, P15 \\
MACS1115JD & $9.2^{+0.4}_{-0.8}$ & $26.2\pm0.2$ & $9.3^{+5.8}_{-3.6}$ & $-18.1 \pm 0.1$ & $17$ & B14 \\
MACS1720JD & $8.9^{+0.3}_{-0.5}$ & $26.9\pm0.3$ & $5.0^{+4.7}_{-0.7}$ & $-18.4^{+0.1}_{-0.2}$ & $0$ & B14 \\
Abel2744JD & $9.8^{+0.2}_{-0.3}$ & $27.37\pm0.16$ & $14.40^{+1.20}_{-1.06}$ & $-18.6 \pm 0.2$ & $0$ & Z14, I15 \\
\hline
\end{tabular}
\caption{A summary of gravitationally-lensed $z > 9$ galaxy candidates. 
References: Z12: \citet{Zheng2012}, B14: \citet{Bouwens2014a}, H15: \citet{Huang2015}, C13: \citet{Coe2013}, GL14: \citet{Gonzalez2014}, P15: \citet{Pirzkal2015}, Z14: \citet{Zitrin2014}, I15: \citet{Ishigaki2015}}
\label{TABLE1}
\end{table*}

\section{Conclusion}
We have performed a large and high resolution cosmological hydrodynamic simulation to investigate the detectability of nebular lines in the rest-frame UV to optical wavelength range at $z > 7$. 
Our new simulation code based on \citet{Okamoto2010} and \citet{Okamoto2014} can reproduce not only the stellar mass function, the downsizing, the mass-metallicity ratio but also the statistical properties of star forming galaxies such as Lyman break galaxies (LBGs), Lyman $\alpha$ emitters and sub-mm galaxies \citep{Shimizu2011, Shimizu2012, Shimizu2014}. 
First, we calibrate our model parameters so as to reproduce not only the stellar mass function at $z \sim 7$ but also the UV luminosity function from $z \sim 7$ to $z \sim 10$. 
Then, we generate a light-cone output which extends from $z = 6$ to $z = 17$ using a number of simulation outputs. 
After making this light-cone output, we adopt the same colour selection criteria the same as in the observations with the HST/WFC3 camera \citep{Bouwens2014a, Oesch2013}. 
Using the colour selected galaxies, we explore expected emission line fluxes in the rest-frame UV to optical wavelength such as C {\sc iv} $1549 \rm \AA$, C {\sc iii}] $1909 \rm \AA$, [O {\sc iii}] $5007 \rm \AA$ and H$\alpha$, etc. 
These lines are very important not only to confirm spectroscopic redshift of the galaxies but also to perform the diagnostics of the gas metallicity. 
We find that the redshift evolution of the line flux as a function of the $H_{160}$ magnitude is fairly weak because the metallicity evolution is very weak as shown in Fig. \ref{SM-METAL_HST}. 
The C {\sc iv} $1549 \rm \AA$ and the C {\sc iii}] $1909 \rm \AA$ lines from galaxies with the band magnitude $ < 26$ AB at $z > 7$ are detectable even by the current telescopes such as the VLT/X-Shooter and the Keck/MOSFIRE. 
Indeed, \citet{Stark2015b} successfully detected the lines from two galaxies at $z \sim 7$ and their detections are fully consistent with our prediction. 
Many important lines are detectable with the JWST/NIRSPEC, JWST/MIRI and the TMT/IRMS if the $H_{160}$ band magnitude is brighter than 28 mag. 
Especially, the C {\sc iv} $1549 \rm \AA$, C {\sc iii}] $1909 \rm \AA$, [O {\sc iii}] $4959/5007 \rm \AA$ and H$\beta$ lines are good targets for them. 

Finally, we predict the detectability of nebular lines for $z \geq 11$ galaxies satisfying K band magnitude $< 30$ mag. 
The redshift evolution of the line flux--K band magnitude relation is very weak even at $z \geq 11$. 
We find that the C {\sc iv} $1549 \rm \AA$ and the C {\sc iii}] $1909 \rm \AA$ lines are detectable by the JWST/NIRSPEC, and TMT/IRMS even at $z \sim 12$. 
However, there are few galaxies at $z \sim 15$ to exceed the detection limit of the future telescopes in our small FOV simulation. 
This does not mean that we cannot detect emission lines from $z \sim 15$ galaxies. 
In fact, the JWST and the TMT can detect emission lines from galaxies with $ < 28$ AB mag at any redshift which would be discovered, 
for example, by a $> 100 ~\rm deg^2$ survey planned with the future telescope such as the WFIRST and the FLARE. 
We can detect $z \sim 15$ galaxies with a help of magnification by gravitational lensing even narrow FOVs. 
We predict that, according to our model, the C {\sc iii}] $1909 \rm \AA$ line in $z > 9$ galaxy candidates (MACS1149JD and MACS0647JD1) is detectable using even the current facilities such as the VLT/X-Shooter and the Keck/MOSFIRE with high probability. 
The survey with the combination of gravitational lensing and the GMT, the TMT, or the JWST is a good way to study higher-$z$ ($z > 7$) line emitting galaxies.

\section*{Acknowledgments}
We thank our referee for his/her constructive comments. 
We would like to thank Tohru Nagao for helpful discussion.
Numerical simulations have been performed with Cray XC30 in CfCA at NAOJ 
and with COMA cluster at Centre for Computational Sciences in University of Tsukuba.  
IS acknowledges the financial support of Grant-in-Aid for Young Scientists (A: 23684007) by MEXT, Japan. 
TO acknowledges the financial support  of Grant-in-Aid for Young Scientists (B: 24740112). 
AKI acknowledges the financial support of Grant-in-Aid for Scientific Research (B: 26287034).

\bsp

\label{lastpage}

\end{document}